\documentclass[11pt]{article}
\usepackage{amssymb,amsmath}
\advance\topmargin-10mm
\advance\textheight20mm
\advance\oddsidemargin-10mm
\advance\textwidth20mm

\begin{document}

\begin{center}
{\Large\bf
An Algorithm of a Virtual Quantum Computer Model on a Classical Computer
}
\bigskip

\textrm{T.\,F. Kamalov, Yu.\,P. Rybakov}
\bigskip

\textit{Physics Department, Moscow State Opened University}

\textit{Department of Theoretical Physics,\\ Russian University of People's Friendship}
\medskip

\textit{E-mail\/}: ykamalov@rambler.ru; qubit@mail.ru
\end{center}

Construction of virtual quantum states became possible due to the hypothesis
on the nature of quantum states quant-ph/0212139. This study considers a
stochastic geometrical background (stochastic gravitational background)
generating correlation (or, coherency) of various quantum non-interacting
objects. In the quantum state virtual model, a simple method of generating
of two (or more) dichotomic signals with controlled mutual correlation
factor from a single continuous stochastic process is implemented. Basing
on the system random number generator of the computer, a model of the
stationary random phase (with the nature of random geometrical background).

Keywords: virtual qubit, virtual quantum computer.

\section{INTRODUCTION}

Simulation of a virtual qubits on a classical computer should not encounter
any insuperable difficulties, as the operating program for quantum states
modelling on a classical computer has been already developed [1]. In the
present study we shall dwell on a classical computer model of quantum
entangled states implemented in Pascal (with Delphy) language. This is an
operating program of a virtual model for two q-bits. It simulates a
controlled correlation or anti-correlation of EPR-Bomm type, including the
total one. As the model considered is a classical one, the Bell's
inequalities are not violated within it. The correlation factor of this
classical model differs from the quantum analog of EPR state by the factor
of $\sqrt{2}$. The simulated operating model of quantum entangled states
and related
discussions simplify understanding of quantum states and EPR-Bohm
correlations. This enables starting the study of quantum algorithms and
programs for quantum computer modelling on a classical computer.

Until now there has not been found any satisfactory way of modelling of quantum
state patterns and their interference on a classical computer. The idea of
quantum calculation put forward by R.Feynman in [2, 3] relies upon the
impossibility of quantum patterns calculation on a classical computer in
virtue of various reasons. At the same time, creation of quantum pattern
models with the help of quantum elements would be rather straightforward. In
its turn, this means that quantum elements must be created. This naturally
means that these elements must interfere with each other, their correlation
factor having to be nonzero. Then, alteration of a single element would
alter the whole quantum pattern. This property of quantum elements is called
quantum parallelism. While the value of an information unit (a bit) in a
classical computer is defined as either 0 or 1, in a quantum computer each
quantum element is described by a wave function $\psi =\alpha \left|
0\right\rangle +\beta \left| 1\right\rangle $,\ meaning this element being
in the state of superposition of zero and unity, $\alpha $,$\beta $\
being
the state complex amplitude (providing $\left| \alpha \right| ^{2}+\left|
\beta \right| ^{2}=1$) with probabilities $P(0)=\left| \alpha \right| ^{2}$,
$P(1)=\left| \beta \right| ^{2}$.

Construction of virtual quantum states became possible due to the hypothesis
on the nature of quantum states [4]. This study considers a stochastic
geometrical background generating correlation (or, coherency) of various
quantum non-interacting objects. The area of this background localization is
called the area of coherence zone. In this area the correlation factor of
various quantum microobjects is nonzero. To explain how this could occur,
let us consider a physical model with the background of stochastic
gravitational fields and waves representing the effect of the stochastic
geometrical background, that is, the physical model with the gravitational
background (i.e. gravitational fields and waves background). This means that
we assume the existence of fluctuations of gravitational waves and fields in
each point of the space, which are mathematically represented by metrics
fluctuations. If we to discuss a possible quantization mechanism, based on
solution concept by Einstein-de Brogle and use the representation of extended
particles as localized self-gravitating structures, we can have the
self-gravitating solution with the non-resonance quantization mechanism as
result [5].

Theoretical investigation of the vacuum accounting for gravitational fields
has been performed in the works by Academician Andrey D.Sakharov [6,7]. In
his first paper of 1967, ''Vacuum Quantum Fluctuations in the Curved Space
and Gravitation Theory'' it has been stated that ''it is assumed in the
modern quantum field theory that the energy-momentum tensor of vacuum
quantum fluctuations equal zero , and the respective action $S(0)$ is
actually zero''. He has further shown that accounting for gravitational
field in the vacuum, taking into consideration definition of space-time
action dependence form curvature in the gravitational theory of A. Einstein
(with invariants of Ricci tensor $R$ and metric tensor $g$), the action
function taking on the form
\[
S(R)=-\frac{1}{16\pi G}\int(dx)\sqrt{-g}R.
\]
Resultant action of all these gravitational fields with number $j$ form the
functional
\[
 S_{0}(\psi)=\sum_{j=1}^{\infty}S_{j},
\]
$\psi (x)$ being the external field given by the metric tensor $g_{ik}$ of
the gravitational field.

Let us consider two classical particles in a field of random gravitational
fields or waves. The General Theory of Relativity gives the length element
in 4-dimensional Riemann space as
\[
  d\ell ^{2}=g_{ik}dx^{i}dx^{k},
\]
the metric in the linear approach is
\[
g_{ik}=\eta_{ik}+h_{ik},
\]
$\eta _{ik}$ being Minkowsky metric, constituting the unity diagonal matrix.
Hereinafter, the indices $i,k,\mu ,\nu ,\gamma ,m,n$ acquire values 0, 1, 2,
3. Indices encountered twice imply summation thereupon. Let us select
harmonic coordinates (the condition of harmonicity of coordinates mean
selection of concomitant frame $\frac{\partial h_{n}^{m}}{\partial x^{m}}=%
\frac{1}{2}\frac{\partial h_{m}^{m}}{\partial x^{n}}$) and let us take into
consideration that $h_{\mu \nu }$ satisfies the gravitational field equations
\[
 \square h_{mn}=-16\pi GS_{mn},
\]
which follow from the General Theory of Relativity; here $S_{mn}$ is
energy-momentum tensor of gravitational field sources with d'Alembertian $%
\square$ and gravity constant $G$. Then, the solution shall acquire the form
\[
h_{\mu\nu}=e_{\mu\nu}\exp(ik_{\gamma}x^{\gamma})+e_{\mu\nu}^{\ast}\exp(ik_{%
\gamma}x^{\gamma}),
\]
where the value $h_{\mu\nu}$\ is called metric perturbation, $e_{\mu\nu}$\
polarization, and $k_{\gamma}$\ is 4-dimensional wave vector. We shall
assume that metric perturbation $h_{\mu\nu}$ are
distributed in space with an unknown distribution function $\rho=\rho
(h_{\mu\nu})$.

Relative displacements $\ell$ of two particles in classic gravitational
fields are described in the General Theory of Relativity by deviation
equations
\[
\frac{D^{2}}{D\tau ^{2}}\ell ^{i}(j)=R_{kmn}^{i}(j)\ell ^{m}\frac{dx^{k}}{%
d\tau }\frac{dx^{n}}{d\tau },
\]
$R_{kmn}^{i}(j)$ being the gravitational field Riemann's tensor with
gravitational field number $j$ of the stohastic gravitational fields.

Specifically, the deviation equations give the equations for two particles
oscillations
\[
\ddot{\ell }^{1}+c^{2}R_{010}^{1}\ell ^{1}=0,\quad \omega =c\sqrt{%
R_{010}^{1}}.
\]

The solution of this equation has the form
\[
\ell ^{1}(j)=\ell _{0}\exp (k_{a}x^{a}+i\omega (j)t),
\]
with $a=1,2,3$. Each gravitational field or wave with index $j$ and
Riemann's tensor $R_{kmn}^{i}(j)$ shall be corresponding to the value $\ell
^{i}(j)$ with random modulated phase $\Phi (j)=\omega (j)t$. If we sum
all fields, we can write $\Phi (t)=\omega (t)t$, where $t$ is the time
coordinate.

This random phase is the same for various quantum microobject in the area of
this coherent background localization[8], this area being defined as the one
within which the correlation factor for these particles is nonzero. Harmonic
oscillations with this type phases could model entangled states.

\begin{center}
2. MODELLING OF A VIRTUAL QUBIT ON A CLASSICAL COMPUTER
\end{center}

In the quantum state virtual model, a simple method of generating of two (or
more) dichotomic random signals with controlled mutual correlation factor
out of a single continuous stochastic process is implemented [1].

Basing on the system random number generator of the computer, a model of the
stationary random process $\Phi (t)$ with $\left\langle \Phi
(t)\right\rangle =0$ has been built determining the random phase evenly
distributed over the interval $0\div 2\pi $. Further, a random signal was
generated on its basis with the help of the algorithm
\[
a(\alpha ,t)=sign\left\{ \cos \left[ \Phi (t)+\alpha \right] \right\} ,
\]

a being $\alpha $ an arbitrary parameter It follows from this definition
that $\left\langle a(\alpha )\right\rangle =0$, $a(\alpha \pm \pi
)=-a(\alpha )$, that is, the signals $a(\alpha )$ and $a(\alpha \pm \pi )$
are anticorrelated. At the same time, $a(\alpha )$ and $a(\alpha \pm \frac{%
\pi }{2})$ are non-correlated signals. In the general case, correlation of
signals $a(\alpha )$ and $a(\alpha +\Delta \alpha )$ is [9]
\[
M(\Delta \alpha )=\left\langle x(\alpha )x(\alpha +\Delta \alpha
)\right\rangle =1-2\left| \Delta \alpha \right| /\pi .
\]

Therefore, out of a single stochastic process $\Phi (t)$ it is possible to
generate two (or more) stochastic dichotomic signals $a(\alpha )$ and $%
a(\alpha +\Delta \alpha )$ with an arbitrary correlation $M(\Delta \alpha )$
between them confined within $-1$ and $+1$. Hence, should the same effect $%
\Phi (t)$ [1] act on several distant observers, then each of the latter
ones, with the help of the ''individual'' local parameter an will be capable
of ''distant influencing'' the paired mutual correlation of observables.
This effect is a classical analog of entangled state's correlation.

\section{INITIALIZATION OF VIRTUAL QUBITS}

Let us consider $N$ virtual qubits constructed according to the above
algorithm. Let us call q-bit No.1 the control, or signal, one.

If the signal qubit is green, then we assign it the value $\left|
0\right\rangle $, if it is red, we output nothing. Hence, at each output we
shall get $N$ initialized integral qubits. In calculations, red color will
correspond to the value $\left| 1\right\rangle $.

\section{HADAMARD TRANSFORM FOR VIRTUAL QUBITS}

Let us assume we have a qubit
\[
\left| q\right\rangle =\frac{1}{\sqrt{2}}(\left| 0\right\rangle +\left|
1\right\rangle );
\]

here, for the state $\left| 0\right\rangle $ the probability $P_{\left|
0\right\rangle }=\frac{1}{2}$, for the state $\left| 1\right\rangle $ the
probability $P_{\left| 1\right\rangle }=\frac{1}{2}$.

However, in the basis rotated by $\frac{\pi }{4}$ the state of the virtual
qubit is determined.

Let the Hadamard transform $H$ be the state of the qubit in the basis rotated
by $\frac{\pi }{4}$. Then,
\[
H\left| 0\right\rangle \rightarrow \frac{1}{2}(\left| 0\right\rangle
+\left| 1\right\rangle )
\]
\[
H\left| 1\right\rangle \rightarrow \frac{1}{2}(\left| 0\right\rangle
+\left| 1\right\rangle ).
\]

Hence, having applied the Hadamard transform to the virtual qubit $q$, we get
the precisely determined value of the qubit
\[
H\left| q\right\rangle =\left| 0\right\rangle  .
\]

\section{LOGICAL COMPONENT $CNOT$ IN A VIRTUAL QUANTUM COMPUTER}

To apply the logical component $CNOT$, using of the above control qubit is
required. The qubit to which the logical operation $CNOT$ is applied will be
called the target qubit. The logical value of the control q-bit is not
altered.

Let us consider two cases:

1. If the control qubit has the value $\left| 1\right\rangle $, then the
target q-bit is switched into the opposite value.

2. If the control qubit has the value $\left| 0\right\rangle $, then the
value of the target q-bit is not altered.

\section{CONCLUSION}

Operation of quantum algorithms on a virtual quantum computer in case of
emulation on a classical computer, a certain time saving should be achieved
in solution of certain problems. However, comparing the computation speed of
real and virtual quantum computers, one could infer, basing on conclusions
of Bell [10] that computation speed of a real quantum computer shall be
higher than that of a virtual quantum computer. This is related to the fact
that the correlation factor for real quantum states, according to Bell's
conclusions, should exceed by the factor of the respective value for
classical models of these quantum states, that is $\sqrt{2}$, for virtual
quantum states. On the other hand, presently there is a lot of problems
related to implementations, measurements, de-coherentization, etc. This
gives rise to the question whether a virtual quantum computer would in the
nearest future possess a gain in time in implementing quantum algorithm for
quantum computation with respect to a real quantum computer. The answer
could be positive due to the host of technical and technological problems
with a real quantum computer impossible to be solved in the near future.
However, even in the contrary case the virtual model will all the same find
an application in classical computers, as it does not require alteration of
latter ones, but rather, extends the capabilities of a classical computer
through special program utilities described above.

\end{document}